# Born again universe

Peter W. Graham,[1] David E. Kaplan,[2] and Surjeet Rajendran[3]

[1]*Stanford Institute for Theoretical Physics, Department of Physics, Stanford University, Stanford, California 94305, USA*
[2]*Department of Physics & Astronomy, The Johns Hopkins University, Baltimore, Maryland 21218, USA*
[3]*Berkeley Center for Theoretical Physics, Department of Physics, University of California, Berkeley, California 94720, USA*

We present a class of nonsingular, bouncing cosmologies that evade singularity theorems through the use of vorticity in compact extra dimensions. The vorticity combats the focusing of geodesics during the contracting phase. The construction requires fluids that violate the null energy condition (NEC) in the compact dimensions, where they can be provided by known stable NEC violating sources such as Casimir energy. The four dimensional effective theory contains an NEC violating fluid of Kaluza-Klein excitations of the higher dimensional metric. These spacetime metrics could potentially allow dynamical relaxation to solve the cosmological constant problem. These ideas can also be used to support traversable Lorentzian wormholes.

## I. INTRODUCTION

Could the Universe have gone through a bounce? This question is of significant import: a bouncing universe could be past eternal and avoid the big bang singularity. It could even potentially permit a new background for cosmological relaxation processes to explain hierarchy problems [1,2], or provide an alternative to the inflationary paradigm [3,4] as the origin of the hot, radiation dominated epoch in our past, see e.g. [5–7]. Reliable implementations of these phenomena require a nonsingular, calculable bounce within semiclassical general relativity, the only known experimentally consistent low energy theory of gravity. Cosmic behavior is constrained by the energy conditions obeyed by matter. For example, it can be shown that a homogeneous, isotropic, positively curved space-time can bounce if the matter violates the strong energy condition [8]. It is of great interest to see if the requirement of positive curvature can be relaxed, permitting a broader range of phenomenological applications. In particular, as we will describe below, a spatially flat bouncing cosmology may permit a solution to the grandest hierarchy problem in physics—namely, the cosmological constant problem.

Singularity theorems would seem to preclude a bounce. It can be shown that a spatially flat, homogeneous and isotropic space-time cannot undergo a nonsingular bounce unless the matter violates the null energy condition (NEC) [8]. There are interesting NEC violating theories that could allow a bounce or other "restart" of the universe (see e.g. [9–11]). However it is difficult to violate the NEC without triggering disastrous short distance instabilities [12]. There are known stable sources that violate the NEC in compact dimensions such as Casimir energy densities and orientifolds [13]. The NEC violation from these sources is an essential ingredient in stabilizing Ricci flat extra dimensions (such as Calabi-Yau compactifications) in an accelerating universe (either during primordial inflation or present day dark energy domination) [14]. But, the NEC violation from these sources is inversely proportional to the volume of the compact space and thus they cannot be used to trigger bounces in a noncompact/large universe.

In this paper, we construct a new class of metrics that permit calculable, nonsingular bouncing cosmologies. The space-time geometry is of the form $R^4 \times X$ where $X$ is a compact manifold with dimensionality $\geq 3$. Importantly, we endow this manifold with a nonfactorizable metric containing vorticity.[1] As we will show, the existence of vorticity enables a nonsingular bouncing cosmology without the need for NEC violation along the noncompact dimensions. In our solutions the NEC is violated, but the violation is along the compact space $X$ and can be provided by stable sources such as Casimir energies. Though we have not constructed the microscopic model of the remaining part of the stress tensor, this part preserves the NEC.

---

[1]The use of vorticity in extra-dimensions to construct nonsingular bouncing cosmologies without the need for NEC violation was considered in [15]. However, the metrics in [15] possess coordinate singularities over which there is a discontinuity in the extrinsic curvature. It can be shown that the localized stress-tensor implied by this discontinuity violates the NEC along the noncompact directions.

In the specific example we study, we will take $X = T^3$, though it should be possible to construct similar solutions for other choices of $X$. We will begin by first presenting the metric in Sec. II and describe the matter necessary to create the bounce. In Sec. III, we show how this metric evades arguments from singularity theorems. The 4D effective theory is discussed in Sec. IV and finally, in Sec. V we comment on possible applications of our ansatz to solve the cosmological constant problem and support traversable wormholes.

## II. THE METRIC ANSATZ

We take our space-time to be of the form $R^4 \times T^3$ with the metric:

$$
\begin{aligned}
ds^2 = &-dt^2 + a(t)^2(dx^2 + dy^2 + dz^2) \\
&+ b^2(d\theta^2 + d\phi_1^2 + d\phi_2^2) \\
&- 2\epsilon b(\sin\theta dt d\phi_1 + \cos\theta dt d\phi_2)
\end{aligned}
\tag{1}
$$

where $a(t)$ is the scale factor for $R^4$ and $b$ the stabilized radius for $T^3$. This metric is isotropic and homogeneous in the spatial $R^3$. Importantly, due to the last two terms in (1), the metric is inhomogeneous along $T^3$ and does not factorize, yielding vorticity. This is the central element that permits our class of solutions. This metric is non-singular for any value of $\epsilon$. Einstein's equations can be used to calculate the stress tensor $T_{\mu\nu}$ necessary to support this metric ansatz. From this, we can identify if the stress tensor needs to violate the NEC and if the required NEC violation exists only along the compact $T^3$ where it could be provided by stable NEC violating sources such as Casimir energies. We present the required stress tensor assuming $\epsilon \ll 1$ only for algebraic convenience, as this is sufficient to highlight the effects of vorticity in (1), and a useful limit for calculating Casimir energy. However, Einstein's equations yield a simple exact expression for the stress tensor and this limit is not necessary for the model's success.

The four-dimensional components of the stress-energy tensor which satisfies Einstein's equations to $\mathcal{O}(\epsilon^2)$ are

$$
T_{tt} = -M_7^5\left(\frac{3\epsilon^2 a''(t)}{a(t)} + \frac{3(\epsilon^2-1)a'(t)^2}{a(t)^2} - \frac{3\epsilon^2}{4b^2}\right)
$$

$$
\begin{aligned}
T_{xx} &= T_{yy} = T_{zz} \\
&= -M_7^5\left(-2(\epsilon^2-1)a(t)a''(t) - (\epsilon^2-1)a'(t)^2 + \frac{\epsilon^2 a(t)^2}{4b^2}\right)
\end{aligned}
\tag{2}
$$

Here $M_7$ is the 7 dimensional Planck scale. Consider a null ray $U^\mu$ along $R^4$. Due to homogeneity and isotropy, we can pick $U^\mu$ to lie along the $x$ direction and check if the NEC is violated along this direction. This yields:

$$
\begin{aligned}
T_{\mu\nu}U^\mu U^\nu &= (U^t)^2\left(T_{tt} - \frac{g_{xx}}{g_{tt}}T_{xx}\right) \\
&= M_7^5\left(\epsilon^2\left(-\frac{a''(t)}{a(t)} - \frac{2a'(t)^2}{a(t)^2} + \frac{1}{2b^2}\right)\right. \\
&\left. - \frac{2(a(t)a''(t) - a'(t)^2)}{a(t)^2}a(t)^2\right)
\end{aligned}
\tag{3}
$$

where we have used the fact that $U^\mu = (U^t, U^x, 0, 0, 0, 0, 0)$ is a null vector. We can now see the effects of the vorticity terms $\propto \epsilon^2$. In their absence, the metric factorizes and yields the standard FRW equations for $R^4 \times T^3$, with $T^3$ stabilized. For a bouncing metric, this leads to a violation of the NEC since at the bounce, $a' = 0$ while $a'' > 0$. The vorticity terms contribute positively to this expression and can thus support a bounce without violating the NEC for any null vector entirely along the noncompact directions $x$, $y$, $z$. For example, if $a(t) = a_0\cosh(\alpha t)$, the vorticity can support a bounce without NEC violation along $U^\mu$ as long as $\epsilon \gg \alpha b$. This implies that the Hubble scale during the bounce is much smaller than the inverse size of the extra-dimensions. It should thus be possible to describe the bounce using a 4D effective theory, see Sec. IV.

For this metric ansatz, there are null rays $V^\mu$ that have components along the extra-dimensions for which the NEC is violated, i.e., $T_{\mu\nu}V^\mu V^\nu < 0$. This NEC violation is independent of the behavior of $a(t)$ but is instead a requirement of the non-zero vorticity in the extra-dimensions. Much like the case of stabilized extra-dimensions in an accelerating universe, we will split the required stress-tensor $(T_{\mu\nu})$ into two pieces: $T_{\mu\nu} = T_C + T_D$. $T_C$ is composed of a suitable choice of Casimir energy densities that will provide a stable NEC violating source along the extra-dimensions. $T_D$ is the additional source necessary to satisfy Einstein's equations. For a given $a(t)$, the form of $T_D$ can be obtained from solving Einstein's equations. We will show that it is possible to find $T_C$ such that $T_D$ preserves the NEC even when the 4D scale factor $a(t)$ undergoes a bounce. We do not have a microscopic field theory that provides $T_D$—thus, we cannot ensure that the source responsible for $T_D$ is devoid of pathologies. However, since $T_D$ respects the NEC, we are optimistic that there might be stable sources of matter that can produce it, and leave this for future work.[2]

---

[2]Moreover, since any NEC preserving fluid can be decomposed into a sum of a fluid that preserves the dominant energy condition (DEC) and a cosmological constant, it might be possible to find classical DEC preserving matter that supports our construction.

In Appendix A, we show that the relevant components of $T_C$, the Casimir part of the stress tensor, can be expressed to $\mathcal{O}(\epsilon^2)$ in the form:

$$T_C = \begin{pmatrix} \rho_1 & 0 & 0 & 0 & 0 & -b\epsilon\sin(\theta)\rho_3 & -b\epsilon\cos(\theta)\rho_3 \\ 0 & -a(t)^2(\rho_1 + \epsilon^2(s_1\rho_1 + s_3\rho_3)) & 0 & 0 & 0 & 0 & 0 \\ 0 & 0 & -a(t)^2(\rho_1 + \epsilon^2(s_1\rho_1 + s_3\rho_3)) & 0 & 0 & 0 & 0 \\ 0 & 0 & 0 & -a(t)^2(\rho_1 + \epsilon^2(s_1\rho_1 + s_3\rho_3)) & 0 & 0 & 0 \\ 0 & 0 & 0 & 0 & b^2\rho_2 & 0 & 0 \\ -b\epsilon\sin(\theta)\rho_3 & 0 & 0 & 0 & 0 & b^2\rho_2 & 0 \\ -b\epsilon\cos(\theta)\rho_3 & 0 & 0 & 0 & 0 & 0 & b^2\rho_2 \end{pmatrix}$$

$$(4)$$

There are additional $\mathcal{O}(\epsilon^2)$ contributions to $T_C$ that are not shown above, for example, to $T_{\phi_1\phi_1}$, $T_{\phi_1\phi_2}$, $T_{\phi_2\phi_2}$. These corrections are subdominant and do not alter our conclusions, as we show in Appendix B. With suitable choice of bosonic and fermionic particles and their masses $M$, we can independently choose $\rho_1$, $\rho_2$, and $\rho_3$.

Exploiting this freedom, we take

$$\rho_2 \sim -\frac{M_7^5}{b^2}, \qquad |\rho_1| \lesssim |\rho_2|, \qquad |\rho_3| \ll |\rho_2|,$$
$$\left(\frac{a'(t)}{a(t)}\right)^2, \qquad \frac{a''(t)}{a(t)} \ll \frac{|\rho_1 + \rho_2|}{M_7^5}. \qquad (5)$$

In Appendix B, we show that this choice ensures that $T_D$ obeys the NEC for every null vector $V^\mu$ as long as the Hubble scale during the bounce is sufficiently small, with $(T_D)_{\mu\nu}V^\mu V^\nu$ strictly positive. The positivity of this quantity implies that the system can support (parametrically) small oscillations of the extra-dimensional moduli without triggering NEC violation. Thus, the stress tensor $T_D$ can also accommodate the stabilization of the compactification. The above parameters are simply a range where the NEC is preserved—it is straightforward to find other such ranges as well. The energy densities necessary for the bounce are sub-Planckian as long as the radius $b$ of the extra-dimension is greater than $1/M_7$, in other words the extra-dimensions are large in Planck units.[3] In our ansatz, the vorticity in the extra-dimensions was assumed to be a constant during the bounce. This assumption was made for convenience: the vorticity can change during the evolution of the universe. As long as it is not zero, we can still find stress tensors $T_D$ that satisfy the NEC during the bounce, exploiting the freedom in choosing $T_C$.

## III. SINGULARITY THEOREMS

Let us see how the metric in (1) avoids singularity theorems [8] that preclude nonsingular bouncing cosmologies. The central element of these theorems is Raychaudhuri's equation which computes the expansion $(\hat{\theta})$ of null (and time-like) congruences. For a null congruence $U^\mu$ parametrized by an affine parameter $\lambda$, we have:

$$\frac{d\hat{\theta}}{d\lambda} = -\frac{1}{2}\hat{\theta}^2 - 2\hat{\sigma}^2 + 2\hat{\omega}^2 - T_{\mu\nu}U^\mu U^\nu \qquad (6)$$

where $\hat{\theta}$, $\hat{\sigma}$ and $\hat{\omega}$ are the expansion, shear and vorticity of the congruence. A nonsingular bounce would require a converging congruence to become diverging. From (6), this is possible only when either $\hat{\omega}$ is nonzero or if the matter violates the NEC.

In our solution, we exploit the fact that the NEC can be violated along the extra dimensions by stable sources such as Casimir energies. This prevents the focusing of null rays that are oriented well inside the extra dimensions. But, this would not prevent the focusing of null congruences that are along the noncompact dimensions $R^4$. The only other way to make these geodesics diverge is to make use of vorticity, i.e., a "centrifugal force" that fights gravitational focusing. This is the case in our nonfactorizable metric: geodesics that are entirely along $R^4$ have nonzero vorticity—they are forced to rotate into the extra-dimensions. This centrifugal force prevents them from converging during a bounce into a singularity, without having to violate the NEC along the noncompact $R^4$. Since we need to prevent the convergence of all 4d null geodesics, we need a compact space where the vorticity can be nonzero everywhere. Vorticity is effectively the curl of a velocity field, and thus we need a compact space $X$ where such a curl can be nonzero everywhere. Topologically, this appears to require at least 3 dimensions. Hence for simplicity we take $X = T^3$, a torus.[4]

---

[3] In Appendix B, we comment on the number of species necessary to satisfy the conditions (5) and their effect on the gravitational cutoff of the theory.

[4] ABC flows are a well-studied example of fluid flow in $T^3$ with nonvanishing vorticity everywhere.

The ability of vorticity to combat the focusing of geodesics is well known. However, its existence is often in conflict with the demand of global hyperbolicity. Vorticity is zero for a congruence that is everywhere hypersurface orthonormal. But, global hyperbolicity is guaranteed as long as we can find a set of surfaces that are orthonormal to a timelike congruence—this congruence need not be a geodesic congruence. Thus global hyperbolicity by itself does not preclude the possibility that geodesics (whose behavior is important for a bounce) possess vorticity.

Nevertheless, the existence of nonvanishing vorticity can cause trouble for global hyperbolicity. A well-known example is that of the Gödel metric where vorticity prevents matter in the universe from undergoing gravitational collapse. But, the rotation results in the existence of closed timelike curves. In our metric, the vorticity moves geodesics into a compact extra-dimension. While the vorticity is nonzero everywhere in the extra-dimensions, its compact size prevents the formation of closed timelike curves. It is straightforward to show that the metric (1) does not possess any closed timelike curves. Consider a curve $\gamma(\tau) = (t(\tau), \ldots)$ in this space-time. If it were to be a closed timelike curve, there must be a point where $\frac{dt}{d\tau}$ is zero. But, at this point (1) is positive definite and the curve is thus spacelike. A similar argument also establishes that causal curves cannot intersect surfaces of constant $t$ more than once. These spacelike surfaces are thus Cauchy surfaces and our space-time is globally hyperbolic.

Every element of our construction is necessary to evade the power of the singularity theorems: NEC violation along compact directions and nonzero vorticity for 4D geodesics evades gravitational focusing, while the restriction of vorticity into just compact dimensions prevents the formation of closed timelike curves enabling the space-time to be globally hyperbolic. Our metric is an example that shows that there are no obstacles in a 7D universe to realize a nonsingular bounce and still retain a homogeneous, isotropic 4D cosmology. The phenomena described by us should thus be a generic feature of metrics with nonvanishing vorticity in the compact space $X$. For example, starting with (1), we can get a large class of bouncing metrics by replacing the $\sin\theta$, $\cos\theta$ terms by $\sin(n\theta)$, $\cos(n\theta)$ where $n \subset Z$. We also present an example where $X = S^3$ in Appendix B.

## IV. EFFECTIVE THEORY

In our scenario, the Hubble scale during the bounce can be parametrically smaller than the compactification scale. In this limit, the entire evolution should be describable in a four-dimensional effective theory. Since there is no vorticity in four dimensions, the singularity theorems imply that the matter that drives the bounce in the four-dimensional description must violate the NEC.

This is indeed the case. As we will see below, the four dimensional description is that of a NEC violating fluid where the fluid consists of a density of Kaluza Klein modes obtained from the decomposition of the higher dimensional metric. Generically, one would expect instabilities in such a fluid. But, these instabilities exist above the mass scale of the particles that compose the fluid. In our case, this mass scale is the compactification scale above which the four-dimensional theory breaks down and the higher dimensional picture becomes relevant. Importantly, the Hubble scale during the bounce can be parametrically ($\propto \epsilon^2$) lower than the compactification scale since the Hubble scale is determined by the density of these massive modes which can be small. In effect, our solution can be viewed as a way to UV complete four dimensional NEC violating matter into a higher dimensional theory, though since the NEC violating matter is parametrically at the same scale as the cutoff of the 4D effective theory, there is not a large range of validity where it would truly be called a 4D, NEC-violating theory.

First, break down the metric into four-dimensional components [16]:

$$g^{(7)}_{AB} = (\det \Phi)^{-1/5} \begin{pmatrix} g_{\mu\nu} + B^a_\mu B^b_\nu \Phi_{ab} & B^c_\mu \Phi_{ca} \\ B^c_\nu \Phi_{cb} & \Phi_{ab} \end{pmatrix} \quad (7)$$

where Greek indices cover the noncompact four dimensions and lower-case latin indices the three compact dimensions. This parametrization is not an approximation, and it is convenient because Einstein's equations are exact to quadratic order in the vector fields $B^a_\mu$. The fields, $g$, $B$, and $\Phi$ can be expanded in Kaluza-Klein (KK) modes along the compact dimensions. We see that the off-diagonal part of the metric is effectively a vacuum expectation value for the vectors $B^{\phi_1}_\mu$, $B^{\phi_2}_\nu$ in the time-direction, specifically, the first-level KK modes along the $\theta$ coordinate.

The four-dimensional components of the Einstein tensor can now be written as:

$$G^{(7)}_{tt} = G_{tt} - \frac{1}{4}((\partial_\theta B^{\phi_1}_t)^2 + (\partial_\theta B^{\phi_2}_t)^2)$$
$$- (B^{\phi_1}_t \partial^2_\theta B^{\phi_1}_t + B^{\phi_2}_t \partial^2_\theta B^{\phi_2}_t) + \cdots \quad (8)$$

$$G^{(7)}_{xx} = G_{xx} - \frac{a(t)^2}{4}((\partial_\theta B^{\phi_1}_t)^2 + (\partial_\theta B^{\phi_2}_t)^2) + \cdots \quad (9)$$

where the ellipses contain, for example, terms with $\Phi$ fields, terms higher order in the vectors $B$, and other terms with derivatives with respect to compact coordinates. The $G_{\mu\nu}$ are components of the four-dimensional Einstein tensor, which contains only the zero-mode ($\theta$-, $\phi_1$-, and $\phi_2$-independent part) of $g_{\mu\nu}$. The other terms shown on the right hand sides can now be considered part of the stress-energy tensor in the four-dimensional effective theory.

The relevant terms in the off-diagonal components of Einstein's equation are

$$-\frac{1}{2}\partial_\theta^2 B_t^{\phi_1} - 3b^2\left(\frac{a'^2}{a^2} + \frac{a''}{a}\right)B_t^{\phi_1} + \cdots = G_{t\phi_1}^{(7)} = \frac{1}{M_7^5}T_{t\phi_1} \tag{10}$$

and similarly for the $t - \phi_2$ component. In the limits we are interested in, the first term dominates. The right-hand side represents a source, that we take as a background field. It, through this equation, is responsible for giving the vector a nonzero expectation value, namely $\langle B_t^{\phi_1}\rangle = (\epsilon/b)\sin\theta$ and $\langle B_t^{\phi_2}\rangle = (\epsilon/b)\cos\theta$.

Thus, we can define

$$M_7^5 T_{tt}^{\text{eff}} \equiv (B_t^{\phi_1}\partial_\theta^2 B_t^{\phi_1} + B_t^{\phi_2}\partial_\theta^2 B_t^{\phi_2})$$
$$+ \frac{1}{4}((\partial_\theta B_t^{\phi_1})^2 + (\partial_\theta B_t^{\phi_2})^2) \to -\frac{3\epsilon^2}{4b^2} \tag{11}$$

$$M_7^5 T_{ii}^{\text{eff}} \equiv \frac{a(t)^2}{4}((\partial_\theta B_t^{\phi_1})^2 + (\partial_\theta B_t^{\phi_2})^2) \to a(t)^2\frac{\epsilon^2}{4b^2} \tag{12}$$

where the index $i = x, y, z$. If one checks the NEC for this effective matter, we see that $n^\mu n^\nu T_{\mu\nu} = -(\epsilon^2/2b^2) < 0$, for a lightlike vector $n^\mu$ pointing any direction in the 4-dimensional subspace. Thus these terms violate the NEC.

Note that the Hubble scale as well as the 4D energy density of the KK modes of the metric can both be parametrically lower than the cutoff of the 4D effective theory. So it would appear that this solution could be described fully in four dimensions, where the fluid can actually be thought of as a dilute condensate of very heavy particles. However the mass of the individual particles making up this fluid is at the cutoff (since they are KK modes). Any potential instability in such a theory lies above the mass scale of the particles, where the 4D effective theory is no longer valid. Thus one interpretation of our model is a UV completion of an NEC-violating four-dimensional theory into extra dimensions, but with no parametric separation between the mass of the NEC-violating 4D matter and the scale where the theory is really seven dimensional.

## V. DISCUSSION

Vorticity in extra-dimensions thus provides a NEC violating source in 4D noncompact space-times. This class of metrics can be used to consistently describe bouncing cosmologies that have been considered in the past (e.g. the Ekpyrotic scenario [17]). It is also straightforward to extend solutions such as the simple harmonic universe [18,19] that avoided cosmological singularity theorems using positively curved spaces into this framework where such solutions can also now describe non-compact geometries. In addition to these possibilities, we discuss two novel applications for these kinds of metrics—relaxation of the cosmological constant and traversable wormholes.

### A. Relaxation of the cosmological constant

Cosmic evolution can allow parameters in the Lagrangian that are not protected by symmetry to naturally relax to fine-tuned values. There are three critical steps in this process. First, there must be a "scanning" mechanism that allows the parameter to change during cosmic history. Second, a "sensor" must be present that realizes that the parameter has attained the desired (fine-tuned) value. Finally, this "sensor" must trigger a backreaction that stops the scanning. It is straightforward to conceive of scanning and backreaction mechanisms. For example, a rolling scalar field provides a natural scanner while the requisite backreaction can be provided by triggering the growth of a potential barrier to prevent further motion of the scalar field. The sensor mechanism however depends on the parameter that is being fine-tuned. In the case of the cosmological relaxation of the gauge hierarchy problem [2], the breaking of electroweak symmetry as the Higgs mass squared goes through zero was used as a sensor to trigger the backreaction that stops the scanning of the electroweak scale.

It is challenging to create a similar sensing mechanism for the cosmological constant, see Abbott's attempt [1]. Any such mechanism has to operate solely through gravity since the cosmological constant does not have any other interactions. But, gravity couples to the entire stress-energy tensor. Thus, a sensing mechanism that is able to trigger a back-reaction to stop the scanning of the cosmological constant can do so only when the energy density in the universe is smaller than the desired fine-tuned value of the cosmological constant. While such a mechanism can be constructed to naturally tune the cosmological constant to small values, the universe thus produced will have an energy density comparable with the observed energy density today, in contradiction with the observational fact that our universe was radiation dominated in the past with temperatures being at least as large as ∼MeV. Thus, as Abbott pointed out, this model has an "empty universe" problem.

Our bouncing cosmology might be able to get around this problem.[5] In Abbott's approach [1], the motion of a scalar field changes the value of the cosmological constant. With technically natural parameters, the scanning can be stopped when the cosmological constant is either slightly positive or negative. Suppose the scanning is stopped when the cosmological constant is slightly negative. The universe will then undergo a crunch. If the universe can go through a calculable, nonsingular bounce it might be possible to reheat the universe to high temperatures while retaining the naturally small value of the cosmological constant.[6]

---

[5]See [20] for an attempt similar in spirit to our approach. Another possibility, see e.g. [21], is to use an NEC-violating fluid to get around the empty-universe problem. This mechanism also has significant challenges to overcome.

This framework circumvents several issues that normally plague attempts to solve the cosmological constant problem. First, the large density of local minima for the scalar field whose motion scans the cosmological constant evades Weinberg's so called "no-go theorem" [22]. This theorem is fundamentally based on the argument that a dynamical solution to the cosmological constant requires two conditions to be simultaneously satisfied—the net potential $V$ of the scanning scalar field must be zero (or close to it) and $\frac{\partial V}{\partial \phi}$ must also be zero so that the field is stabilized in a minimum (or near it). For a generic potential, while it is possible to engineer one of these two conditions, they cannot both be satisfied without fine-tuning. In the Abbott potential, where there is a large density of minima, the second condition is easy to satisfy. It is thus sufficient to engineer dynamics that can satisfy the first condition, namely, making the net cosmological constant/potential energy small.

A second issue arises if one tries to solve the cosmological constant problem when the universe is at high temperatures. At high temperatures (e.g. prior to the QCD phase transition), the cosmological constant is expected to have been significantly higher than it is today. Any mechanism that operates at this time would have to rapidly adjust the value of the cosmological constant in the short time between the QCD phase transition and the time of recombination. This is hard since the cosmological constant is a small contribution to the energy density of the universe at these early times and if it only couples gravitationally, it is difficult to see how any such mechanism could operate. In our scenario, the tuning of the cosmological constant occurs when the energy densities in the universe are comparable to the mean density of the universe today—i.e., the standard model is in its vacuum and thus the tuned value of the cosmological constant will correspond to the vacuum value and not some high temperature value. As the universe undergoes a bounce and becomes hot, the cosmological constant will change as expected during a phase transition. But, during this time, since the universe is radiation dominated, the effects of the large cosmological constant are not observable (as in our own universe). The subsequent expansion and cooling of the universe will move the cosmological constant back to its vacuum value, where the scalar field has appropriately tuned it to be small.

Thus, the "empty universe" problem of Abbott is actually a virtue since it results in tuning the cosmological constant in the correct vacuum. A bounce that reheats the universe could complete the Abbott model and solve the cosmological constant problem.[7] We leave further study of this possibility for future work.

### B. Wormholes

The behavior of geodesics during a bounce wherein a converging congruence has to become diverging is also shared by wormhole geometries. Since vorticity can prevent the focusing of geodesics in space-times filled with NEC preserving matter, it is interesting to ask if a construction similar to (1) can yield traversable wormholes.[8] We present an example that realizes the Morris-Thorne wormhole [23] which connects two asymptotically flat spatial regions together:

$$ds^2 = -dt^2 + dr^2 + (l^2 + r^2)d\theta^2 + (l^2 + r^2)\sin^2\theta d\phi^2 \\ + b^2(d\psi_1^2 + d\psi_2^2 + d\psi_3^2) \\ + 2\epsilon b(\sin\psi_1 dt d\psi_2 + \cos\psi_1 dt d\psi_2) \quad (13)$$

Here $(t, r, \theta, \phi)$ are the coordinates on $R^4$, $(\psi_1, \psi_2, \psi_3)$ are the coordinates on $T^3$ and $l$ parametrizes the size of the wormhole's throat. Much like the analysis for the bounce, we can now use Einstein's equations to compute the stress tensor necessary to support this wormhole and check if it violates the NEC. For null geodesics $U^\mu$ that are entirely along $R^4$, we find:

$$T_{\mu\nu}U^\mu U^\nu = \frac{\epsilon^2(l^2 + r^2)^2 - 4b^2 l^2}{2b^2(l^2 + r^2)^2} \quad (14)$$

In the absence of vorticity, the stress tensor necessary to support (13) violates the NEC, while in its presence the NEC need not be violated along $R^4$. Similar to the case of the bounce, there are null-vectors $V^\mu$ that are pointed into the extra-dimensions for which the NEC is violated. But, one may again provide the required NEC violation in the extra-dimensions using stable sources such as Casimir energy densities. We leave further exploration of these matters for future work.

### VI. CONCLUSIONS

We have shown that qualitatively new phenomena are possible in 7D space-times of the form $R^4 \times X$ where the metric is nonfactorizable and possesses vorticity in the extra-dimensions. These phenomena arise from the fact that vorticity prevents the focusing of null congruences, a central need for the construction of nonsingular bounces and traversable wormholes. These metrics require the presence of a NEC preserving source whose microphysics we have not identified. It would be interesting to see if there are restrictions from particle physics on such sources when they carry vorticity. If so, it could lead to a new class of conditions on viable stress-energy tensors where the dynamic behavior of the source is restricted, and a new class of singularity theorems. On the other hand, if viable sources are found, they could greatly enlarge the phenomenology of general relativity.

---

[6] A universe thus born again might potentially be free of the sin of fine-tuning.

[7] This harmonious marriage of inflation and ekpyrosis could thus potentially allow for concurrent solutions to both the gauge hierarchy and cosmological constant problems through dynamic relaxation.

[8] The potential use of extra-dimensions to produce wormholes was first considered in the film *Interstellar*.

## ACKNOWLEDGMENTS

We are grateful to Koushik Balasubramanian for patiently answering our questions and for not charging us his current consulting fee. We thank Victor Gorbenko and Mehrdad Mirbabayi for discussions regarding Casimir energies. S. R. would like to thank Sandip Trivedi for early discussions on these topics. We thank Nima Arkani-Hamed, Savas Dimopoulos, Robert Jaffe, Ted Jacobsen, Shamit Kachru, Nemanja Kaloper and Raman Sundrum for discussions. S. R. was supported in part by the NSF under Grants No. PHY-1638509 and No. PHY-1507160, the Alfred P. Sloan Foundation Grant No. FG-2016-6193 and the Simons Foundation Award 378243. P. W. G. acknowledges the support of NSF Grant No. PHY-1720397 and DOE Early Career Award DE-SC0012012. D. E. K. acknowledges the support of NSF Grant No. PHY-1214000. This work was supported in part by Heising-Simons Foundation Grants No. 2015-037 and No. 2015-038.

## APPENDIX A: CASIMIR CALCULATIONS

The stress energy tensor necessary to create the metric in (1) violates the null energy condition (NEC) for geodesics that extend into $T^3$. This violation is dominantly due to the existence of vorticity in the extra-dimensions and is essentially independent of the bouncing behavior of the 4D scale factor $a(t)$. For the purpose of an existence proof, we show below that with suitable choice of parameters, the stress tensor of Casimir energies in the $T^3$ can provide a safe NEC violating source to support (1).

We work in the limit $a(t) = a_0$, a constant. To see that this is sufficient, observe that in the limit $\epsilon \to 0$, the metric factorizes and thus the Casimir energies in $T^3$ contribute like a cosmological constant along 4D, their assumed form in (4). Corrections to the Casimir stress tensor arising from vorticity and a time-dependent scale factor are parametrically of order $\epsilon a''/a, \epsilon a'/a$. Since we work in the limit $a''/a, (a'/a)^2 \ll \epsilon^2/b^2$, these corrections are parametrically smaller than the required NEC violation $\sim (\epsilon/b^2)$. Rescaling away $a_0$, our problem reduces to finding the Casimir energies for $R^4 \times T^3$ with the metric:

$$ds^2 = -dt^2 + (dx^2 + dy^2 + dz^2) + b^2(d\theta^2 + d\phi_1^2 + d\phi_2^2) - 2\epsilon b(\sin\theta dt d\phi_1 + \cos\theta dt d\phi_2). \tag{A1}$$

Clearly, in the limit $\epsilon \ll 1$, the vorticity terms can be treated as a perturbation over a flat Minkowski background. I.e., $g_{\mu\nu} = \eta_{\mu\nu} + \epsilon h_{\mu\nu}$ where $\eta_{\mu\nu}$ is the flat Minkowski metric obtained from $g_{\mu\nu}$ in the limit $\epsilon \to 0$. The Casimir energies for a field $\Psi$ can be computed by treating the perturbation $\epsilon h_{\mu\nu}$ as a background insertion.

To $\mathcal{O}(\epsilon^2)$, the Casimir energy for $\Psi$ takes the form:

$$\begin{pmatrix} \rho_1 & 0 & 0 & 0 & 0 & -\epsilon\sin(\theta)\rho_3 & -\epsilon\cos(\theta)\rho_3 \\ 0 & -\rho_1 + \epsilon^2\tilde\rho_1 & 0 & 0 & 0 & 0 & 0 \\ 0 & 0 & -\rho_1 + \epsilon^2\tilde\rho_1 & 0 & 0 & 0 & 0 \\ 0 & 0 & 0 & -\rho_1 + \epsilon^2\tilde\rho_1 & 0 & 0 & 0 \\ 0 & 0 & 0 & 0 & (\rho_2 + \epsilon^2\tilde\rho_5) & 0 & 0 \\ -\epsilon\sin(\theta)\rho_3 & 0 & 0 & 0 & 0 & (\rho_2 + \epsilon^2\tilde\rho_6) & \epsilon^2\tilde\rho_{67} \\ -\epsilon\cos(\theta)\rho_3 & 0 & 0 & 0 & 0 & \epsilon^2\tilde\rho_{67} & (\rho_2 + \epsilon^2\tilde\rho_7) \end{pmatrix} \tag{A2}$$

As explained in Appendix B, the terms $\tilde\rho_5, \tilde\rho_6, \tilde\rho_7, \tilde\rho_{36}, \tilde\rho_{37}$ and $\tilde\rho_{67}$ do not affect the proof that the vorticity inducing stress tensor $T_D$ preserves the null energy condition at $\mathcal{O}(\epsilon^2)$. Thus, we will not compute them.

For a given field, $\rho_1, \tilde\rho_1, \rho_2$ and $\rho_3$ are functions of the mass $M$ of the field and the radius $b$. Our computations below show that $\tilde\rho_1$ is a linear combination of $\rho_1$ and $\rho_3$, while $\rho_1, \rho_2$ and $\rho_3$ are algebraically independent, i.e., there is no fixed function $f$ such that $\rho_i = f(\rho_j, \rho_k)$. This is not a surprise—the algebraic independence of $\rho_1$ and $\rho_2$ for a massive scalar field is well established, while $\rho_3$ depends upon the specific metric perturbation[9] Thus, while they are all functions of $M$ (and $b$), by picking various masses and exploiting the relative minus sign between fermions and bosons (for periodic boundary conditions), we can find a set of bosons and fermions whose net Casimir energy takes the form (5).

We now establish the algebraic independence of $\rho_1$, $\rho_2$ and $\rho_3$ for a scalar field $\Psi$ of mass $M$. To $\mathcal{O}(\epsilon^2)$, the Lagrangian is

$$\mathcal{L} \supset -\frac{1}{2}\sqrt{-g}\Bigg(\eta^{\mu\nu}\partial_\mu\Psi\partial_\nu\Psi + M^2\Psi^2 + \underbrace{\epsilon^2(\partial_t\Psi)^2}_{O_1}$$
$$-\frac{2\epsilon}{b}\Big(\underbrace{\partial_t\Psi\partial_{\phi_1}\Psi\sin\theta}_{O_2} + \underbrace{\partial_t\Psi\partial_{\phi_2}\Psi\cos\theta}_{O_3}\Big)$$
$$+\underbrace{-\frac{\epsilon^2}{b^2}(\sin\theta\partial_{\phi_1}\Psi + \cos\theta\partial_{\phi_2}\Psi)^2}_{O_4}\Bigg). \tag{A3}$$

The stress energy tensor for this field is calculated from the Green's function through the relation $T_{\mu\nu} = \lim_{x_2\to x_1}\overline{\partial_\mu\partial_\nu}\,G(x_2, x_1)$, where the partial derivatives are taken to be symmetrized on the $x_1$, $x_2$ coordinates as $\overline{\partial_\mu\partial_\nu} \equiv (\partial_\mu^2\partial_\nu^1 + \partial_\mu^1\partial_\nu^2)/2$. This $T_{\mu\nu}$ contains a divergent Poincaré invariant contribution to the cosmological constant—upon subtracting this piece, the remainder is the Casimir energy. The boundary conditions necessary to obtain $G(x_2, x_1)$ in $R^4 \times T^3$ can be incorporated by first starting with the corresponding Green's function $G^\infty(x_2, x_1)$ in $R^7$ and using the method of images to enforce periodic boundary conditions along the $\theta$, $\phi_1$ and $\phi_2$ directions: $G(x_2, x_1) = \sum_{\vec{n}} G^\infty(x_2, x_1 + 2\pi\vec{n})$ where $\vec{n} = l\hat{\theta} + m\hat{\phi}_1 + n\hat{\phi}_2$, $(l, m, n)$ are integers, and $(\hat{\theta}, \hat{\phi}_1, \hat{\phi}_2)$ are unit vectors along $\theta$, $\phi_1$, and $\phi_2$. Corrections to $G^\infty(x_2, x_1)$ from the operators in (A3) can be calculated perturbatively. The operators $O_4$ preserve 4D Lorentz invariance—they can thus at most contribute to renormalizations of $\rho_1$ at $\mathcal{O}(\epsilon^2)$ and are thus irrelevant. $O_1$, $O_2$ and $O_3$ in (A3) are the leading contributions to $\rho_1$, $\rho_2$, $\rho_3$ and $\tilde{\rho}_1$ through the Feynman diagrams in Fig. 1.

The Casimir stress tensor is

$$T_{\mu\nu} = \lim_{x_2\to x_1}\underbrace{\Bigg(\overline{\partial_\mu\partial_\nu}\sum_{\vec{n}} G^\infty(x_2, x_1 + 2\pi\vec{n})\Bigg)}_{S}$$
$$- \lim_{x_2\to x_1}(\overline{\partial_\mu\partial_\nu}G^\infty(x_2, x_1)). \tag{A4}$$

This difference is independent of the regulator used to compute the divergent contributions to the cosmological constant. The integrals over the momenta $p^{\theta, \phi_1, \phi_2}$ in the sum (labeled S) in (A4) can be performed using the Poisson summation formula, reducing these integrals to sums over $\vec{n}$. The resulting expressions can be regulated using dimensional regularization for the integrals and Zeta function regularization [24] for the sums. As a check on our calculations, we also calculated the relevant components of the Casimir tensor numerically. This avoids the use of any regulators for the sums or integrals as the CC term ($\vec{n} = 0$) in the above sum is the only term that is actually divergent and it is subtracted off. The numeric results agreed with our analytic expressions given below to many digits.

The perturbed propagators (Fig. 1) are

$$G_0^\infty(x_2, x_1) \propto i\int\frac{d^7p}{(2\pi)^7}\frac{e^{ip\cdot(x_2-x_1)}}{p^2 + M^2} \tag{A5}$$

$$G_{1a}^\infty(x_2, x_1) \propto -i\epsilon\int\frac{d^7p}{(2\pi)^7}\frac{e^{ip\cdot(x_2-x_1)}(ip^tp^{\phi_1})}{p^2 + M^2}$$
$$\times\left(\frac{e^{-ix_1^\theta}}{p_+^2 + M^2} - \frac{e^{ix_1^\theta}}{p_-^2 + M^2}\right) \tag{A6}$$

$$G_{1b}^\infty(x_2, x_1) \propto -i\epsilon\int\frac{d^7p}{(2\pi)^7}\frac{e^{ip\cdot(x_2-x_1)}(p^tp^{\phi_2})}{p^2 + M^2}$$
$$\times\left(\frac{e^{-ix_1^\theta}}{p_+^2 + M^2} + \frac{e^{ix_1^\theta}}{p_-^2 + M^2}\right) \tag{A7}$$

$$G_{2a}^\infty(x_2, x_1) \propto i\epsilon^2\int\frac{d^7p}{(2\pi)^7}\frac{e^{ip\cdot(x_2-x_1)}p^{t2}}{(p^2 + M^2)^2} \tag{A8}$$

$$G_{2b+2c}^\infty(x_2, x_1) \propto -i\epsilon^2\int\frac{d^7p}{(2\pi)^7}\frac{e^{ip\cdot(x_2-x_1)}p^{t2}}{p^2 + M^2}$$
$$\times\Bigg(\frac{p^{\phi_12} + p^{\phi_22}}{p^2 + M^2}\left(\frac{1}{p_-^2 + M^2} + \frac{1}{p_+^2 + M^2}\right)$$
$$+ \frac{(p^{\phi_22} - p^{\phi_12})2\cos(x_1^\theta + x_2^\theta)}{(p_-^2 + M^2)(p_+^2 + M^2)}\Bigg) \tag{A9}$$

---

[9]For example, it can be shown that for metrics where the $\sin\theta$, $\cos\theta$ terms are replaced by $\sin(n\theta)$, $\cos(n\theta)$ for $n \subset Z$, $\rho_3$ scales as $1/n^2$, while $\rho_1$, $\rho_2$ are unaltered. The conditions (5) are easily satisfied in this case.

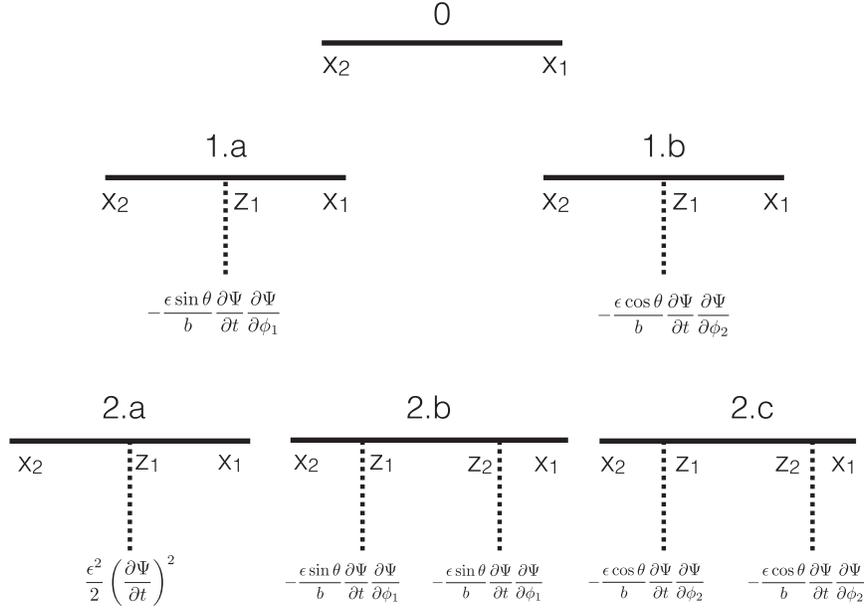

FIG. 1. Corrections to the propagator $G^\infty(x_2, x_1)$ from the perturbations $O_1$, $O_2$ and $O_3$. There is also a diagram 2.d which is similar to 2.b and 2.c except that it has one $\sin\theta$ vertex and one $\cos\theta$ vertex and they are summed in both orders.

$$G_{2d}^\infty(x_2, x_1) \propto -i\epsilon^2 \int \frac{d^7 p}{(2\pi)^7} \frac{e^{ip\cdot(x_2-x_1)} p'^2 p^{\phi_1} p^{\phi_2} \sin(x_1^\theta + x_2^\theta)}{(p^2+M^2)(p_-^2+M^2)(p_+^2+M^2)} \tag{A10}$$

where for example $G_{1a}^\infty$ corresponds to the correction to the propagator from the diagram labeled "1(a)" in Fig. 1. We define $p_\pm^2 = -p'^2 + \vec{p}\cdot\vec{p} + (p^\theta \pm 1/2)^2 + p^{\phi_1 2} + p^{\phi_2 2}$ where $p^\mu$ is the momentum along the $\mu$ direction. Note that $G_{2d}$ and the second term in $G_{2b+2c}$ do contribute to the Casimir tensor but their contributions do not need to be calculated exactly because they are not relevant for our purposes (e.g. it can be easily shown that $G_{2d}$ contributes only to $\tilde{\rho}_{67}$). Moreover, since our argument relies on the functional independence of the terms $\rho_1$, $\rho_2$ and $\rho_3$ we did not compute overall numerical prefactors for the above Green's functions.

The leading order contributions to $\rho_1$, $\rho_2$, $\rho_3$ and $\tilde{\rho}_1$ are

$$\rho_1 \propto -\frac{45 e^{-2\pi b M}}{32\pi^7 b^7} - \frac{45 M e^{-2\pi b M}}{16\pi^6 b^6} - \frac{9 M^2 e^{-2\pi b M}}{4\pi^5 b^5} - \frac{3 M^3 e^{-2\pi b M}}{4\pi^4 b^4} + \text{higher order exponentials}$$

$$\rho_2 \propto -\frac{15 e^{-2\pi b M}}{8\pi^7 b^7} - \frac{15 M e^{-2\pi b M}}{4\pi^6 b^6} - \frac{27 M^2 e^{-2\pi b M}}{8\pi^5 b^5} - \frac{7 M^3 e^{-2\pi b M}}{4\pi^4 b^4} - \frac{M^4 e^{-2\pi b M}}{2\pi^3 b^3} + \text{higher order exponentials}$$

$$\rho_3 \propto -\frac{1}{64\pi^6 b^7} \int_0^1 da (\pi y_1^2 K_2(2\pi\sqrt{y_1}) + \pi y_2^2 K_2(2\pi\sqrt{y_2}) + 2y_1^{3/2} K_3(2\pi\sqrt{y_1}) + 2y_2^{3/2} K_3(2\pi\sqrt{y_2}))$$

$$+ \text{rapidly converging exponentials and Bessels}$$

where $y_1 = (1 + b^2 M^2 - a)$ and $y_2 = (b^2 M^2 + a)$,

$$\tilde{\rho}_1 = s_1 \rho_1 + s_3 \rho_3. \tag{A11}$$

One can show that the coefficients $s_1$ and $s_3$ are nonzero, but their exact values do not matter for our analysis. All of these expressions can be numerically evaluated for various values of $M$, establishing their algebraic independence. Using the fact that there is a relative minus sign between the Casimir energies for bosons and fermions, it is possible to find a suitable combination of fermions and bosons to obtain the desired values of $\rho_1$, $\rho_2$ and $\rho_3$.

## APPENDIX B: THE NULL ENERGY CONDITION

In this section, we prove that with the choice of parameters in (5), the stress tensor $T_D$ obeys the NEC. We work in the limit $(a'/a)^2$, $a''/a \ll \epsilon^2/b^2$. To prove that $T_D$ obeys the NEC, we need to show that $(T_D)_{\mu\nu} V^\mu V^\nu > 0$ for every null vector $V^\mu$, where the vector can be located at any point in $R^4 \times T^3$. Without loss in generality, we may choose $V^\mu = (1, \beta_x, 0, 0, \beta_\theta, \beta_{\phi_1}, \beta_{\phi_2})$. Fix $\beta_x$ by the condition $g_{\mu\nu}V^\mu V^\nu = 0$. For any point on $R^4 \times T^3$, $(T_D)_{\mu\nu}V^\mu V^\nu$ is now a function $(\beta_\theta, \beta_{\phi_1}, \beta_{\phi_2})$. Compute the Hessian of this function and observe that the eigenvalues are of the form $-(\rho_1 + \rho_2) + \mathcal{O}(\epsilon^2)$. In the limit of small $\epsilon$, by taking $(\rho_1 + \rho_2) < 0$, we can ensure that all the eigenvalues of the Hessian are positive. This implies that

$(T_D)_{\mu\nu}V^\mu V^\nu$ is a convex function of $(\beta_\theta, \beta_{\phi_1}, \beta_{\phi_2})$ at any point on $R^4 \times T^3$. Minimize $(T_D)_{\mu\nu}V^\mu V^\nu$ as a function of $(\beta_\theta, \beta_{\phi_1}, \beta_{\phi_2})$. Since $(T_D)_{\mu\nu}V^\mu V^\nu$ is convex, this minimum is the global minimum. We demand that for any point on $R^4 \times T^3$, this minimum is positive, ensuring that the NEC is preserved everywhere.

To $\mathcal{O}(\epsilon^2)$, this minimum is

$$\frac{\epsilon^2(4b^4(\rho_1^2(s_1+1) + \rho_1(\rho_2 s_1 + \rho_3(s_3 + 2)) + \rho_3(\rho_3 + \rho_2 s_3)) - 2b^2 M_7^5(\rho_1 - \rho_2 + 2\rho_3) + M_7^{10})}{4b^4(\rho_1 + \rho_2)}. \tag{B1}$$

To make this minimum positive, start by taking $\rho_2 < 0$ and $|\rho_1| < |\rho_2|$ so that $\rho_1 + \rho_2 < 0$ (ensuring that we are at a minimum). Choosing $\rho_3 \cong 0$ and $\rho_1 = \delta \rho_2$, the minima are of the form:

$$(T_D)_{\mu\nu}V^\mu V^\nu|_{\min} = \frac{\epsilon^2(-4b^4 \delta s_1 \rho_2^2 + 2b^2(2\delta + 1)M_7^5 \rho_2 + (\delta + 1)M_7^{10})}{4b^4 \rho_2} + \mathcal{O}(\delta^2) \tag{B2}$$

It is clear that by choice of the sign of $\delta$ and making $\rho_2$ sufficiently negative, the above expression can be made positive for any given nonzero values of $s_1$ and $s_3$. This analysis also makes it clear that terms such as $\tilde{\rho}_5$ in (A2) are irrelevant to this calculation—they do not appear in (B1). The most dangerous terms for the NEC check are terms where $\beta_x$ is large. These null rays do not extend much into the extra-dimension where the NEC violating Casimir energies have most of their impact. For these rays, $\tilde{\rho}_1$ matters since the contribution from the $\rho_1$ terms in (A2) are effectively canceled. This is not the case for terms such as $\tilde{\rho}_5$—the null rays that would cancel contributions from $\rho_2$ benefit from the NEC violating Casimir and are not dangerous.

The tuning in (5) requires densities $\rho_2 \sim M_7^5/b^2$, whereas the typical Casimir energy densities for a single species are $\sim \kappa/b^7$ [see (A11)] where $\kappa$ is the 7D loop factor. The number of species required to accomplish this goal is $N \sim (bM_7)^5/\kappa$. This large number of species will renormalize the Planck scale, lowering the cutoff $\Lambda$ of the theory below the 7D Planck scale $M_7$ (although this could be avoided at the cost of tuning the Planck scale). Requiring that the natural value for the renormalized Planck scale $\sim \kappa N \Lambda^5$ not be larger than $M_7$ implies that $(b\Lambda) \sim \mathcal{O}(1)$. Thus, in this model, parametric separation between the compactification scale $1/b$ and the cutoff $\Lambda$ of the theory cannot be achieved without tuning in the gravitational sector. In the absence of tuning, one might worry that these large energy densities that are at the cutoff of the theory may invalidate our classical analysis, for example, by sourcing higher dimensional operators in the gravitational action. But, even though these Casimir energy densities are large, in our solution, they are canceled to high accuracy by $T_D$, leading to curvature scales $\propto \epsilon^2/b^2 \ll \Lambda^2$. Since the curvature is small, the higher dimensional gravitational corrections are not relevant.

Casimir energies are an inefficient way to violate the NEC. It is thus of interest to investigate other sources that can more efficiently violate the NEC in compact directions. This includes examples such as orientifolds and compact space-times with positive curvature. For example, since positive curvature positively contributes as a NEC violating source in Einstein's equations, it can be used to significantly decrease the need for additional NEC violation. Consider the space-time $R^4 \times S^3$ with the metric:

$$ds^2 = -dt^2 + a(t)^2(dx^2 + dy^2 + dz^2)$$
$$+ b^2(d\Psi^2 + \sin^2\Psi d\theta^2 + \sin^2\Psi \sin^2\theta d\phi^2) \tag{B3}$$

$$-2\epsilon b(\cos\theta dt d\Psi - \sin\Psi \cos\Psi \sin\theta dt d\theta - \sin^2\Psi \sin^2\theta dt d\phi) \tag{B4}$$

where the coordinates on $R^4$ are $(t, x, y, z)$ while $(\Psi, \theta, \phi)$ are the coordinates on $S^3$. This metric possesses vorticity and similarly allows a bounce without the need for NEC violating sources along $R^4$. NEC violation is necessary along the extra-dimensions—but the amount necessary is $\sim a''/a$, parametrically smaller than the violation necessary in the case of $R^4 \times T^3$. This significantly reduced NEC violation could potentially be provided by Casimir energies, which would require far fewer species than the case considered above. Although we have not explicitly calculated the Casimir energies for this geometry, we expect the different components of the Casimir stress tensor to be sufficiently independent to allow this construction. The detailed calculation is left for future work.

[1] L. F. Abbott, A mechanism for reducing the value of the cosmological constant, Phys. Lett. 150B, 427 (1985).

[2] P. W. Graham, D. E. Kaplan, and S. Rajendran, Cosmological Relaxation of the Electroweak Scale, Phys. Rev. Lett. 115, 221801 (2015).

[3] A. D. Linde, A new inflationary universe scenario: A possible solution of the horizon, flatness, homogeneity, isotropy and primordial monopole problems, Phys. Lett. 108B, 389 (1982).

[4] A. H. Guth, The inflationary universe: A possible solution to the horizon and flatness problems, Phys. Rev. D 23, 347 (1981).

[5] P. J. Steinhardt and N. Turok, A cyclic model of the universe, Science 296, 1436 (2002).

[6] P. Creminelli and L. Senatore, A smooth bouncing cosmology with scale invariant spectrum, J. Cosmol. Astropart. Phys. 11 (2007) 010.

[7] D. Battefeld and P. Peter, A critical review of classical bouncing cosmologies, Phys. Rep. 571, 1 (2015).

[8] S. W. Hawking and G. F. R. Ellis, *The large scale structure of space-time* (Cambridge University Press, Cambridge, 1973).

[9] P. Creminelli, M. A. Luty, A. Nicolis, and L. Senatore, Starting the universe: Stable violation of the null energy condition and non-standard cosmologies, J. High Energy Phys. 12 (2006) 080.

[10] V. A. Rubakov, Phantom without UV pathology, Teor. Mat. Fiz. 149, 409 (2006) [Theor. Math. Phys. 149, 1651 (2006)].

[11] P. Creminelli, A. Nicolis, and E. Trincherini, Galilean genesis: An alternative to inflation, J. Cosmol. Astropart. Phys. 11 (2010) 021.

[12] V. A. Rubakov, The null energy condition and its violation, Usp. Fiz. Nauk 184, 137 (2014) [Phys. Usp. 57, 128 (2014)].

[13] E. G. Gimon and J. Polchinski, Consistency conditions for orientifolds and d manifolds, Phys. Rev. D 54, 1667 (1996).

[14] P. J. Steinhardt and D. Wesley, Dark energy, inflation and extra dimensions, Phys. Rev. D 79, 104026 (2009).

[15] K. Balasubramanian and S. P. Dabholkar, Time-dependent warping and non-singular bouncing cosmologies, arXiv:1401.7015.

[16] T. Appelquist and A. Chodos, The quantum dynamics of Kaluza-Klein theories, Phys. Rev. D 28, 772 (1983).

[17] J. Khoury, B. A. Ovrut, P. J. Steinhardt, and N. Turok, The ekpyrotic universe: Colliding branes and the origin of the hot big bang, Phys. Rev. D 64, 123522 (2001).

[18] P. W. Graham, B. Horn, S. Kachru, S. Rajendran, and G. Torroba, A simple harmonic universe, J. High Energy Phys. 02 (2014) 029.

[19] P. W. Graham, B. Horn, S. Rajendran, and G. Torroba, Exploring eternal stability with the simple harmonic universe, J. High Energy Phys. 08 (2014) 163.

[20] P. J. Steinhardt and N. Turok, Why the cosmological constant is small and positive, Science 312, 1180 (2006).

[21] L. Alberte, P. Creminelli, A. Khmelnitsky, D. Pirtskhalava, and E. Trincherini, Relaxing the cosmological constant: A proof of concept, J. High Energy Phys. 12 (2016) 022.

[22] S. Weinberg, The cosmological constant problem, Rev. Mod. Phys. 61, 1 (1989).

[23] M. S. Morris, K. S. Thorne, and U. Yurtsever, Wormholes, Time Machines, and the Weak Energy Condition, Phys. Rev. Lett. 61, 1446 (1988).

[24] E. Elizalde, Zeta function methods and quantum fluctuations, J. Phys. A 41, 304040 (2008).